\documentclass[conference,a4paper]{APSIPA2021}
\usepackage{amsmath}
\usepackage[pdftex]{graphicx}
\usepackage{multirow}
\usepackage{algorithm}
\usepackage{algpseudocode}
\usepackage{threeparttable}
\usepackage{amsmath, amssymb, bm}

\usepackage{geometry}
\usepackage{fancyhdr}

\fancypagestyle{firststyle}{
  \fancyhf{}
}

\begin{document}

\title{A Linear-Time Algorithm for the Closest Vector Problem of Triangular Lattices} 

\author{
\authorblockN{
Kenta Takahashi\authorrefmark{1} and
Wataru Nakamura\authorrefmark{2}
}

\authorblockA{
\authorrefmark{1}
Hitachi, Ltd., Japan \\
E-mail: kenta.takahashi.bw@hitachi.com}

\authorblockA{
\authorrefmark{2}
Hitachi, Ltd., Japan \\
E-mail: wataru.nakamura.va@hitachi.com}
}

\maketitle
\thispagestyle{firststyle}
\pagestyle{fancy}

\begin{abstract}
  Fuzzy Extractor (FE) and Fuzzy Signature (FS) are useful schemes for generating cryptographic keys from fuzzy data such as biometric features. Several techniques have been proposed to implement FE and FS for fuzzy data in an Euclidean space, such as facial feature vectors, that use triangular lattice-based error correction. In these techniques, solving the closest vector problem (CVP) in a high dimensional (e.g., 128--512 dim.) lattice is required at the time of key reproduction or signing. However, solving CVP becomes computationally hard as the dimension $n$ increases. In this paper, we first propose a CVP algorithm in triangular lattices with $O(n \log n)$-time whereas the conventional one requires $O(n^2)$-time. Then we further improve it and construct an $O(n)$-time algorithm.
\end{abstract}

\section{Introduction}
Fuzzy Extractor (FE) \cite{Dodis2008} is a general method to extract a (reproducible) random string from fuzzy data which may include errors within a certain range. FE enables the realization of cryptographic systems using biometric information or PUFs (Physically Unclonable Functions) \cite{Maes2013} as key management mediums. Similarly, Fuzzy Signature (FS) \cite{Takahashi2019} is a digital signature scheme that uses fuzzy data itself as a signing key.

The specific algorithms of FE and FS depend on the metric space to which the fuzzy data belong and the error range to be tolerated. For example, if the fuzzy data is in $n$-bit Hamming space and you want to tolerate up to $t$-bit errors, known FE algorithms are using $n$-bit error correcting codes that can correct $t$-bit errors \cite{Dodis2008}. For the strength of the keys to be extracted, the error correcting code (ECC) should have as many codewords as possible for given $(n,t)$. In this sense, it is ideal for the ECC to be a {\it perfect code} \footnote{Although there are a limited number of $(n,t)$ pairs for which a perfect code exists.}, which divides the $n$-bit space completely with Hamming spheres of radius $t$.

On the other hand, in biometric authentication such as face recognition, features are often extracted as $n$-dimensional vectors and errors are often evaluated as Euclidean distance.
For example, feature extraction methods based on deep metric learning such as CosFace \cite{Wang2018} and ArcFace \cite{Deng2019}, where a feature vector is $L_2$-normalized and the 'closeness' between two vectors $\bm{x}, \bm{y}$ is defined by the cosine similarity $S_C(\bm{x}, \bm{y}) := \frac{\bm{x} \cdot \bm{y}}{\|\bm{x}\| \|\bm{y}\|} = \bm{x} \cdot \bm{y}$ are often used for face recognition. This is equivalent to using Euclidean distance, since $\|\bm{x} - \bm{y}\|^2 = 2(1 - S_C(\bm{x}, \bm{y}))$. 
Applying FE and FS here requires an error correction mechanism in Euclidean space, but since $n$-dimensional space cannot be divided by $n$-dimensional hyperspheres, an approximate division must be used instead. As a technique for this purpose, a method using an $n$-dimensional lattice structure has been proposed \cite{Yoneyama2015, Zhang2021, Katsumata2021}. The desirable properties of the lattice to be applied to FE and FS are as follows.
\begin{enumerate}
    \item The lattice gives a dense sphere packing.
    \item The closest lattice vector to a given arbitrary vector can be computed efficiently.
\end{enumerate}
The first property is important to ensure the strength of the key to be extracted. Such lattice structures have been studied for many years in the context of the sphere packing problem \cite{Conway1993}, but for the densest structures it is an open problem except in $n = 1,2,3,8$ and $24$ dimensions. The second requirement is also important to optimize the balance between security and performance. However, the CVP in general lattices is known to be NP-hard \cite{Dinur2003}.
As a lattice structure that satisfies these properties to some extent, Yoneyama et al. proposed the use of triangular lattices and constructed an algorithm for CVP with $O(n^2)$-time complexity for $n$-dimensions by exploiting the high symmetry of the lattice \cite{Yoneyama2015}. However, when applied to biometric systems, this algorithm may take too much time to compute. For example, the typical number of feature dimensions in face recognition is around $n = 128$ -- $512$, and when applied to a 1:N authentication system with $N=100,000$, the evaluated processing time is 3.8 -- 62 seconds, as shown in Sec.~\ref{Sec:experiment}.

In this paper, we propose fast CVP algorithms for triangular lattices with complexity of $O(n \log n)$-time and $O(n)$-time. The main ideas are as follows. The first is to speed up the coordinate transformation subroutine, which conventionally took $O(n^2)$-time, to $O(n)$-time by using a kind of memorization technique that exploits the properties of the ``good'' representation of the basis matrix of the triangular lattice. The second is to speed up the subroutine that searches for the closest vector from $n$ candidates, which conventionally took $O(n^2)$-time, to $O(n)$-time by showing that the sequence of distances from the target vector to the suitably sorted candidates is convex and introducing a binary search. These two ideas yield an $O(n \log n)$-time CVP algorithm. Further, by carefully introducing an $O(n)$-time selection algorithm instead of an $O(n \log n)$-time sorting algorithm, we obtain an $O(n)$-time CVP algorithm.
This paper is an advanced version of the work presented at APSIPA ASC 2024 \cite{Takahashi2024}.
\section{Preliminaries and Related Works}

\subsection{Triangular Lattice}
In an $n$-dimensional real Euclidean space $\mathbb{R}^n$, take $n$ independent vectors $\bm{b}^1,\cdots,\bm{b}^n \in \mathbb{R}^n$  (basis vectors) and let $B=(\bm{b}^1,\cdots,\bm{b}^n)$ be a matrix (basis matrix). Then the lattice $L(B)$ is defined as
\[
L(B):=\{ B \boldsymbol{\bm{x}} ~|~ \bm{x} \in \mathbb{Z}^n \}.
\]
That is, $L(B)$ is the set of all vectors $\bm{b}^1,\cdots,\bm{b}^n$ linearly combined with integer coefficients.

A lattice $L(B)$ is called a triangular lattice if it has a basis $B=(\bm{b}^1,\cdots,\bm{b}^n)$ satisfying the following conditions.
\[
\forall_i \| \bm{b}^i \| =c ~\text{(constant)}, ~ \forall_{i \ne j} \bm{b}^i \cdot \bm{b}^j=c^2/2.
\]

\begin{figure}
    \centering
    \includegraphics[width=0.8\linewidth]{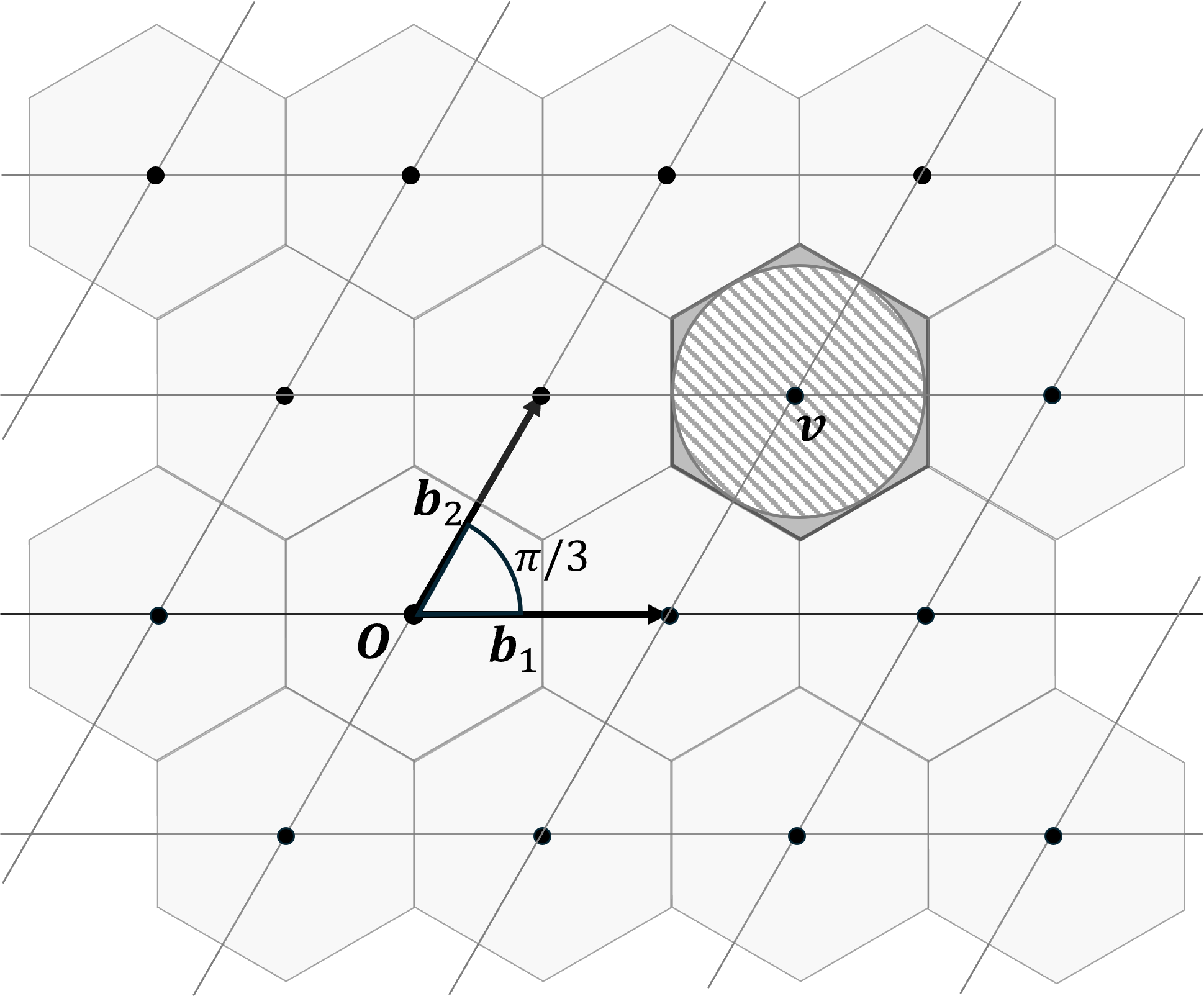}
    \caption{Black dots are the points of a triangular lattice in $\mathbb{R}^2$ spanned by $B=(\bm{b}^1,\bm{b}^2 )$. The gray hexagon represents a Voronoi cell centered at lattice vector $\bm{v}$. The shaded disk represents a sphere with center $\bm{v}$ whose radius is half the minimum distance.}
    \label{fig:TriLattice}
\end{figure}

The above conditions imply that an angle between any two basis vectors is $\pi/3$ (Fig.\ref{fig:TriLattice}). Triangular lattices have been proven to give the densest sphere-packing in two and three dimensions. It has also been shown experimentally to have relatively good performance in higher dimensions when applied to FE \cite{Yoneyama2015} and FS \cite{Katsumata2021}.
For simplicity, and without loss of generality, the following discussion deals with the triangular lattice with $c=1$ and a ``good'' basis matrix represented \footnote{There are many basis representations of the triangular lattice. Among them we use a ``good'' representation to construct efficient coordinate transformation algorithms.} as
\begin{equation}
    B = \begin{pmatrix}
    \alpha_1 & \beta_1 & \beta_1 & \cdots & \beta_1 \\
    0 & \alpha_2 & \beta_2 & \cdots & \beta_2 \\
    0 & 0 & \alpha_3 & \cdots & \beta_3 \\
    \vdots & \vdots & \vdots & \ddots & \vdots \\
    0 & 0 & 0 & \cdots & \alpha_n
    \end{pmatrix},
    \label{eq:BasisMat}
\end{equation}
where $\alpha_k$ and $\beta_k$ are constants computed by the following simultaneous difference equations:
\[
\alpha_k = \sqrt{1 - \sum_{i=1}^{k-1} \beta_i^2}, \quad
\beta_k = \frac{\frac{1}{2} - \sum_{i=1}^{k-1} \beta_i^2}{\alpha_k} \quad (k \in [n]).
\]
Here $[n]$ denotes the set $\{1,\cdots,n\}$. 
The first few values are as follows:
\[
\begin{aligned}
&\alpha_1 = 1, \quad \alpha_2 = \frac{\sqrt{3}}{2}, \quad \alpha_3 = \frac{\sqrt{6}}{3}, \quad \cdots , \\
&\beta_1 = \frac{1}{2}, \quad \beta_2 = \frac{\sqrt{3}}{6}, \quad \beta_3 = \frac{\sqrt{6}}{12}, \quad \cdots.
\end{aligned}
\]

\subsection{Conventional CVP Algorithm for Triangular Lattices}
A CVP algorithm takes as input a target vector $\bm{y} \in \mathbb{R}^n$ and outputs the closest lattice vector. In the following, we outline the conventional CVP algorithm for the triangular lattice proposed by Yoneyama et al. See reference \cite{Yoneyama2015} for details.

For the sake of simplicity in the following discussion, we will assume that the ``decimal parts'' $r_1, \cdots r_n$ of $w_1, \cdots, w_n$ are all different values. 
However, the algorithms proposed in this paper can be extended to the case where some of the values are the same.

\begin{algorithm}
\caption{Conventional CVP Algorithm  $CV(\bm{y})$}
\begin{algorithmic}[1]
\Require $\bm{y} \in \mathbb{R}^n$
\State $(x_1, \cdots, x_n)^t \gets B^{-1} \bm{y}$
\For{$i \gets 1, \cdots, n$}
    \State $w_i \gets \lfloor x_i \rfloor$
    \State $r_i \gets x_i - w_i$
\EndFor
\State $\bm{w} \gets (w_1, \cdots, w_n)^t$
\State $(r_{\sigma(1)}, \cdots, r_{\sigma(n)}) \gets \text{Sort}(r_1, \cdots, r_n)$ \Comment{descending order}
\For{$k \gets 0, \cdots, n$}
    \State $\bm{z}^k \gets (z_1^k, \cdots, z_n^k)^t$ where
    \[
    z_j^k \gets
    \begin{cases} 
    1, & \text{for } j = \sigma(1), \cdots, \sigma(k) \\
    0, & \text{for } j = \sigma(k+1), \cdots, \sigma(n)
    \end{cases}
    \]
\EndFor
\State $\bm{z} \gets \arg \min_{\bm{z}^k \in \{\bm{z}^0, \cdots, \bm{z}^n\}} \|B \bm{r} - B \bm{z}^k\|$
\State \Return $B(\bm{w} + \bm{z})$
\end{algorithmic}
\end{algorithm}

The time complexity of each step is $O(n^2)$ for step 1, 8--10 and 12, $O(n)$ for step 2--6 and $O(n \log n)$ for step 7. Step 11 appears to take $O(n^3)$-time at first glance, but can be computed in $O(n^2)$-time by using the following difference equation:
\[
B \bm{z}^n = 0, \quad
B \bm{z}^{k-1} = B \bm{z}^k + b_{\sigma(k)} \quad (\text{for } k = n, n-1, \cdots, 1).
\]
Therefore, the total complexity is $O(n^2)$.

\section{A Quasilinear-time Algorithm}
Based on the conventional algorithm, let us consider speeding up steps 1, 8--10 and 12, which take $O(n^2 )$-time. 
Specifically, the coordinate transformation process (product of basis matrix and vector) in steps 1 and 12 and the closest vector search (enumeration of candidate vectors and distance calculation) in steps 8--10 are accelerated as subroutines, respectively.

\subsection{Coordinate Transformation Subroutine}
\label{subsec:CTS}

Let us consider algorithms for mutual transformation between 
$\bm{y} = (y_1, \cdots, y_n)^t$ (in Cartesian coordinate system) and 
$\bm{x} = (x_1, \cdots, x_n)^t$ (in triangular lattice coordinate system) 
that satisfies the following relation:

\[
\bm{y} = Bx \Leftrightarrow \bm{x} = B^{-1}\bm{y}.
\]

In general, the product of a matrix and a vector requires $O(n^2)$-time, but by utilizing the basis matrix representation in Eq.~\eqref{eq:BasisMat}, a linear-time 
algorithm can be constructed as follows. Defining $s_k \ (k \in [n])$ as
\[
s_k := \sum_{i=k+1}^n x_i,
\]
we have
\begin{equation}
    s_n = 0, ~~ s_{k-1} = s_k + x_k ~ (k = 2, \cdots, n). 
    \label{eq:sk}
\end{equation}
Furthermore, from Eq.~\eqref{eq:BasisMat} the following equation holds:
\begin{equation}
    y_k = \alpha_k x_k + \beta_k s_k \quad (k = 1, \cdots, n).
    \label{eq:yk}
\end{equation}
Therefore, given $(x_1, \cdots, x_n)$, $(y_1, \cdots, y_n)$ can be obtained by solving Eq.~\eqref{eq:sk}~\eqref{eq:yk} recursively in the following order: 
\[
y_n \to s_{n-1} \to y_{n-1} \to s_{n-2} \to \cdots \to s_1 \to y_1.
\]
Similarly, given $(y_1, \cdots, y_n)$, we can obtain $(x_1, \cdots, x_n)$ by solving Eq.~\eqref{eq:sk}~\eqref{eq:yk} recursively in the following order:
\[
x_n \to s_{n-1} \to x_{n-1} \to s_{n-2} \to \cdots \to s_1 \to x_1.
\]
The time complexity of these algorithms is $O(n)$.

\begin{algorithm}
\caption{Triangular to Cartesian Coordinates $T2C(\bm{x})$}
\begin{algorithmic}[1]
\Require $\bm{x} = (x_1, \cdots, x_n)^t \in \mathbb{R}^n$
\State $s_n \gets 0$
\For{$k \gets n, n-1, \cdots, 1$}
    \State $y_k \gets \alpha_k x_k + \beta_k s_k$
    \State $s_{k-1} \gets s_k + x_k$
\EndFor
\State \Return $\bm{y} = (y_1, \cdots, y_n)^t$
\end{algorithmic}
\end{algorithm}

\begin{algorithm}
\caption{Cartesian to Triangular Coordinates $C2T(\bm{y})$}
\begin{algorithmic}[1]
\Require $\bm{y} = (y_1, \cdots, y_n)^t \in \mathbb{R}^n$
\State $s_n \gets 0$
\For{$k \gets n, n-1, \cdots, 1$}
    \State $x_k \gets \frac{y_k - \beta_k s_k}{\alpha_k}$
    \State $s_{k-1} \gets s_k + x_k$
\EndFor
\State \Return $\bm{x} = (x_1, \cdots, x_n)^t$
\end{algorithmic}
\end{algorithm}

\subsection{Closest Vector Search Subroutine}
\label{subsec:CVSS}

Here we describe an idea to speed up steps 8--11 of the conventional CVP algorithm. The goal here is to efficiently determine the $k$ that minimize 
\[
d_k := \|B \bm{r} - B \bm{z}^k\|^2
\]
without calculating $\bm{z}^k$ and $d_k$ for each $k = 0, 1, \cdots, n$. $d_k$ can be expanded as follows:

\begin{align*}
    d_k =& \|B \bm{r}\|^2 + \|B \bm{z}^k\|^2 - 2 B \bm{r} \cdot B \bm{z}^k \\
        =& \|B \bm{r}\|^2 + \|\bm{b}^{\sigma(1)} + \cdots + \bm{b}^{\sigma(k)}\|^2 \\
         & - 2(r_1 \bm{b}^1 + \cdots + r_n \bm{b}^n) \cdot (\bm{b}^{\sigma(1)} + \cdots + \bm{b}^{\sigma(k)}) \\
        =& \|B \bm{r}\|^2 + \frac{1}{2}k(k+1) \\
         & - k(r_1 + \cdots + r_n) - (r_{\sigma(1)} + \cdots + r_{\sigma(k)}) \\
        =& \|B \bm{r}\|^2 + \frac{1}{2}k^2 + k\left(\frac{1}{2} - r_{\text{sum}}\right) - \sum_{i=1}^k r_{\sigma(i)},
\end{align*}
where $r_{\text{sum}} := \sum_{i=1}^n r_i$. Therefore the first and second-order differences of $d_k$ are:

\begin{align*}
    \Delta_k    &:= d_k - d_{k-1} = k - r_{\text{sum}} - r_{\sigma(k)}, \\
    \Delta_k^2  &:= \Delta_k - \Delta_{k-1} = 1 - (r_{\sigma(k)} - r_{\sigma(k-1)}).
\end{align*}

Since $0 \leq r_i < 1 \ (i \in [n])$, we have $\Delta_k^2 > 0$. Therefore, if we regard $d_k$ as a (discrete) function of $k$, it is convex downwards, and the $k$ that gives the minimum value is:

\begin{enumerate}
    \item[(i)] if $\Delta_1 \geq 0$ then $k = 0$,
    \item[(ii)] if $\Delta_n \leq 0$ then $k = n$,
    \item[(iii)] otherwise the $k \in [n-1]$ satisfying $\Delta_k \leq 0 \leq \Delta_{k+1}$ \quad $\Leftrightarrow r_{\sigma(k)} \geq k - r_{\text{sum}} \geq r_{\sigma(k+1)} - 1$.
\end{enumerate}

\subsection{Algorithm Description}
Algorithm \ref{Alg:QLinCV} is the proposed CVP algorithm $QLinCV$ that uses the subroutine described in Sec.~\ref{subsec:CTS} and Sec.~\ref{subsec:CVSS} to speed up the conventional one.
The asymptotic complexity of $QLinCV$ is $O(n \log n)$-time.

\begin{algorithm}
\caption{Quasilinear-time CVP Algorithm $QLinCV(\bm{y})$}
\label{Alg:QLinCV}
\begin{algorithmic}[1]
\Require $\bm{y} \in \mathbb{R}^n$
\State $(x_1, \cdots, x_n)^t \gets C2T(\bm{y})$
\For{$i \gets 1, \cdots, n$}
    \State $w_i \gets \lfloor x_i \rfloor$
    \State $r_i \gets x_i - w_i$
\EndFor
\State $\bm{w} \gets (w_1, \cdots, w_n)^t$
\State $(r_{\sigma(1)}, \cdots, r_{\sigma(n)}) \gets \text{Sort}(r_1, \cdots, r_n)$ \Comment{descending order}
\State $r_{\text{sum}} \gets \sum_{i=1}^n r_i$
\If{$1 - r_{\text{sum}} - r_{\sigma(1)} \geq 0$}
    \State $k \gets 0$
\ElsIf{$n - r_{\text{sum}} - r_{\sigma(n)} \leq 0$}
    \State $k \gets n$
\Else
    \State Determine $k \in [n-1]$ satisfying the following inequality by a binary search.
    \[
    k - r_{\sigma(k)} \leq r_{\text{sum}} \leq k + 1 - r_{\sigma(k+1)}
    \]
\EndIf
\State $\bm{z} \gets (z_1, \cdots, z_n)^t$ where
\[
z_j \gets
\begin{cases}
1, & \text{for } j = \sigma(1), \cdots, \sigma(k) \\
0, & \text{for } j = \sigma(k+1), \cdots, \sigma(n)
\end{cases}
\]
\State \Return $T2C(\bm{w} + \bm{z})$
\end{algorithmic}
\end{algorithm}

\section{A Linear-time Algorithm}
Here, we aim to further speed up the algorithm from the previous section, that is, to achieve linear time.
The dominant part in the computational complexity of Algorithm~\ref{Alg:QLinCV} was the sorting (step 7).
Then, is it possible to determine $k$ which minimizes $d_k$ and calculate $\bm{z}=\bm{z}^k$ in $O(n)$-time without sorting $(r_1,\cdots,r_n)$?
The answer is yes. The following explains this idea.

\subsection{Determination of $k$}
\label{subsec:Detk}
Firstly, let's take a closer look at the condition in (iii) of Sec.~\ref{subsec:CVSS}, i.e., $\Delta_k \le 0 \le \Delta_{k+1}$.
If we let $\hat{k} := \lfloor r_{sum} \rfloor$ then $\Delta_{\hat{k}} \le 0$ holds. 
Furthermore, if $\hat{k} \le n-2$ then $\Delta_{\hat{k}+2} > 0$ holds. 
Therefore the $k$ satisfying the condition of (iii) is either $k=\hat{k}$ or $k=\hat{k}+1$.
(Note that since $\Delta_n>0$ holds in the condition of (iii), if $\hat{k}=n-1$ then $k=\hat{k}=n-1$.)

From the above discussions the $k$ that gives the minimum of $d_k$ can be determined if we can calculate $\Delta_i$ for $i \in \{1, \hat{k}, \hat{k}+1, n \}$. 
Since $\Delta_i = i-r_{sum}-r_{\sigma(i)}$ and depends on $\sigma(i)$, it may seem that the $Sort(r_1,\cdots,r_n)$ step is essential in the algorithm.
Nevertheless, the sorting is actually not necessary because we do not have to know the whole sorted values $r_{\sigma(1)}, \cdots, r_{\sigma(n)}$, but only need to know the up to four values $\{ r_{\sigma(1)}, r_{\sigma(\hat{k})}, r_{\sigma(\hat{k}+1)}, r_{\sigma(n)} \}$. 
Each $r_{\sigma(i)}$ is the $i$-th largest value among $(r_1, \cdots, r_n)$ and can be determined in $O(n)$-time using efficient selection algorithm such as \cite{Musser1997}. 
Thus each $\Delta_i$ is calculated in  $O(n)$-time, and the $k$ is also determined in $O(n)$-time without sorting.

\subsection{Calculation of $z_k$}
\label{subsec:Calczk}
Next, let's consider how to calculate step 16 in Algorithm~\ref{Alg:QLinCV} without referring the sorting result $\sigma(1), \cdots, \sigma(n)$, but only using $r_{\sigma(k)}$.
If you note that $i \le k \Leftrightarrow r_{\sigma(i)} \ge r_{\sigma(k)}$, you can see that $j \in \{\sigma(1),\cdots,\sigma(k)\} \Leftrightarrow r_j \ge r_{\sigma(k)}$.
Therefore, each $z_j$ in step 16 can be determined by just comparing each $r_j$ with $r_{\sigma(k)}$ in $O(1)$-time, and thus $\bm{z}=(z_1,\cdots,z_n)$ can be calculated in $O(n)$-time without sorting.

\subsection{Algorithm Description}
Algorithm \ref{Alg:LinCV} is the proposed linear-time CVP algorithm $LinCV$ that uses the ideas described in Sec.\ref{subsec:Detk} and Sec.\ref{subsec:Calczk} to speed up the Algorithm~\ref{Alg:QLinCV}.
The asymptotic complexity of $LinCV$ is $O(n)$-time.

\begin{algorithm}
\caption{Linear-time CVP Algorithm $LinCV(\bm{y})$}
\label{Alg:LinCV}
\begin{algorithmic}[1]
\Require $\bm{y} \in \mathbb{R}^n$
\State $(x_1, \cdots, x_n)^t \gets C2T(\bm{y})$
\For{$i \gets 1, \cdots, n$}
    \State $w_i \gets \lfloor x_i \rfloor$
    \State $r_i \gets x_i - w_i$
\EndFor
\State $\bm{w} \gets (w_1, \cdots, w_n)^t$
\State $r_{\text{sum}} \gets \sum_{i=1}^n r_i$
\State $\hat{k} \gets \lfloor r_{\text{sum}} \rfloor$
\State Determine the $i$-th largest $r_{\sigma(i)}$ for $i \in \{ 1, \hat{k}, \hat{k} + 1, n \}$ using a linear time selection algorithm.
\If{$1 - r_{\text{sum}} - r_{\sigma(1)} \geq 0$}
    \State $k \gets 0$
\ElsIf{$n - r_{\text{sum}} - r_{\sigma(n)} \leq 0$}
    \State $k \gets n$
\ElsIf{$r_{\text{sum}} \leq \hat{k} + 1 - r_{\sigma(\hat{k}+1)}$}
    \State $k \gets \hat{k}$
\Else
    \State $k \gets \hat{k} + 1$
\EndIf
\State $\bm{z} \gets (z_1, \cdots, z_n)^t$ where
\[
z_j \gets
\begin{cases}
1, & \text{if } r_j \geq r_{\sigma(k)} \\
0, & \text{otherwise}
\end{cases}
\]
\State \Return $T2C(\bm{w} + \bm{z})$
\end{algorithmic}
\end{algorithm}

\section{Experimental Evaluation}
\label{Sec:experiment}

\begin{figure}
    \centering
    \includegraphics[width=1.0\linewidth]{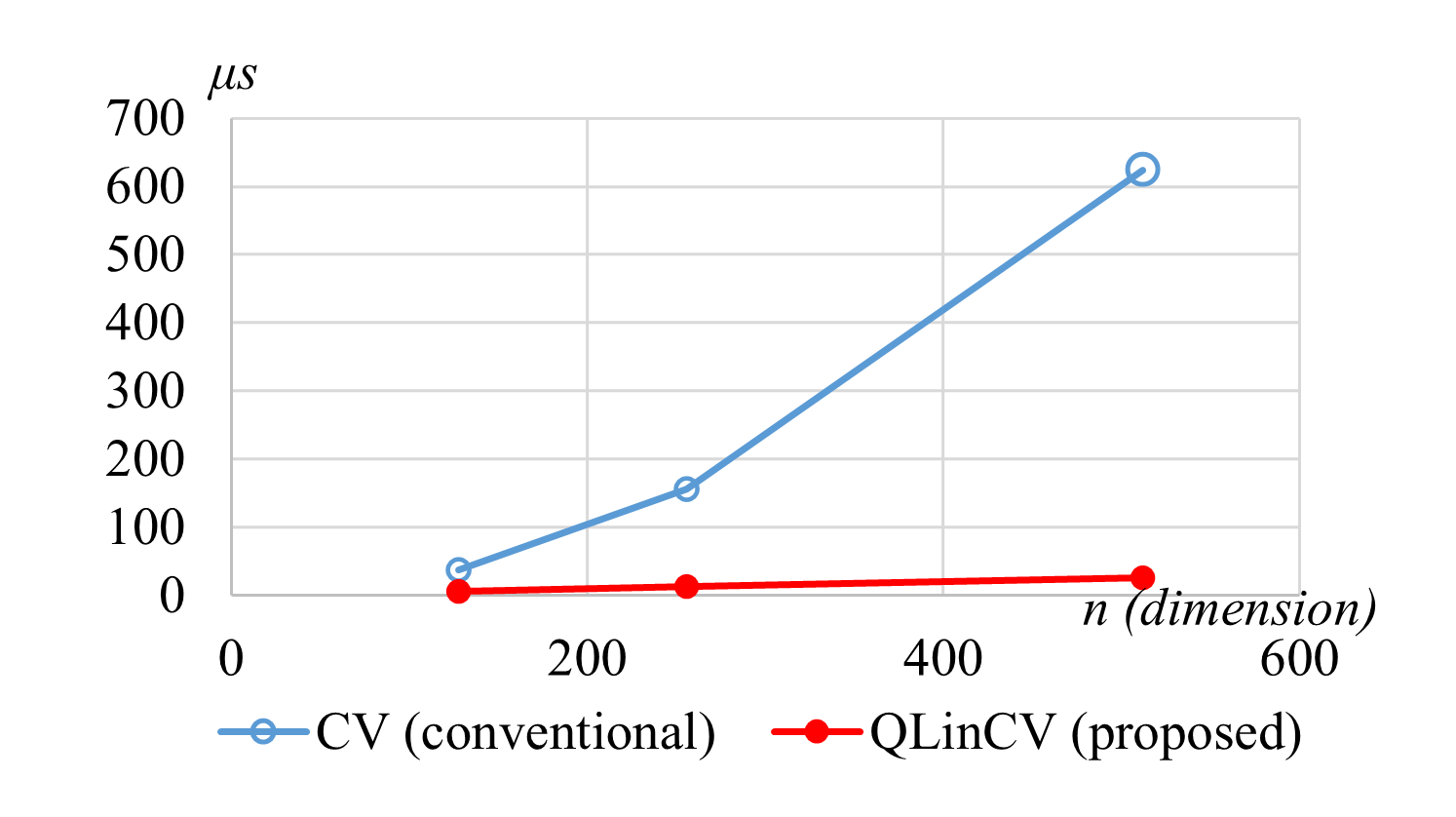}
    \caption{Graphical comparison of the computation time of the algorithms.}
    \label{fig:Experiment}
\end{figure}

We implemented the proposed CVP algorithm $QLinCV$ and the conventional algorithm $CV$ to evaluate their computation times experimentally. The implementation and evaluation of $LinCV$ is a future work.
We randomly selected 10,000 real vectors $y_1, \cdots, y_{10,000}$ from the range $y_i \in [-100, 100]^n$ as target vectors, executed $CV(y_i)$ and $QLinCV(y_i)$ for all $y_i$ ($i = 1, \cdots, 10,000$), measured the total computation time for each algorithm, and evaluated the average time per run for each algorithm. The results for $n = 128, 256,$ and $512$ are shown in Tab.\ref{tbl:Experiment} and Fig.\ref{fig:Experiment}. The computing environment is as follows: OS: Windows 10 Pro, CPU: Intel(R) Core(TM) i7-8700K @ 3.70GHz, memory 32GB, storage: SSD.

$QLinCV$ is about 6.6 times faster than $CV$ when $n = 128$, about 12 times faster when $n = 256$, and about 24 times faster when $n = 512$. The number of processes per second when $n = 512$ is $3.82 \times 10^4$ with $QLinCV$, while it is $1.60 \times 10^3$ with $CV$.

\begin{table}[h!]
\centering
\caption{Average Computation time of the algorithms ($\mu$s)}
\label{tbl:Experiment}
\begin{tabular}{|c|c|c|c|}
\hline
Algorithm & $n = 128$ & $n = 256$ & $n = 512$ \\ \hline \hline
CV (conventional) & 38.2 & 156 & 623 \\ \hline
QLinCV (proposed) & 5.8 & 12.5 & 26.2 \\ \hline \hline
Speedup factor     & 6.6 & 12   & 24 \\ \hline
\end{tabular}
\end{table}

\section{Conclusions}
In this paper, we proposed fast algorithms $QLinCV$ and $LinCV$ for the closest vector problem (CVP) on high-dimensional triangular lattices. The time complexity of $QLinCV$ and $LinCV$ are $O(n \log n)$ and $O(n)$ respectively for $n$ dimensions, whereas that of conventional algorithms $CV$ is $O(n^2)$. Experimental evaluation shows that $QLinCV$ achieves a speedup of about 24 times compared with $CV$ for $n = 512$ dimensions. 

The CVP algorithm for high-dimensional triangular lattices is useful for constructing Fuzzy Extractors (FE) and Fuzzy Signatures (FS) using biometric data such as facial feature vectors. Therefore, it is expected to realize real-time biometric identification for large-scale users with biometric template protection \cite{ISO24745} based on FE and FS such as the Public Biometric Infrastructure (PBI) \cite{Takahashi2019}. A more detailed evaluation (e.g., using large-scale face datasets) is a subject for future work.


\end{document}